# Deterministic Annealing Based Optimization for Zero-Delay Source-Channel Coding in Networks

Mustafa S. Mehmetoglu, *Student Member, IEEE,* Emrah Akyol, *Member, IEEE,* Kenneth Rose, *Fellow, IEEE*

*Abstract*—This paper studies the problem of global optimization of zero-delay source-channel codes that map between the source space and the channel space, under a given transmission power constraint and for the mean square error distortion. Particularly, we focus on two well known network settings: the Wyner-Ziv setting where only a decoder has access to side information and the distributed setting where independent encoders transmit over independent channels to a central decoder. Prior work derived the necessary conditions for optimality of the encoder and decoder mappings, along with a greedy optimization algorithm that imposes these conditions iteratively, in conjunction with the heuristic noisy channel relaxation method to mitigate poor local minima. While noisy channel relaxation is arguably effective in simple settings, it fails to provide accurate global optimization in more complicated settings considered in this paper. We propose a powerful non-convex optimization method based on the concept of deterministic annealing – which is derived from information theoretic principles and was successfully employed in several problems including vector quantization, classification and regression. We present comparative numerical results that show strict superiority of the proposed method over greedy optimization methods as well as prior approaches in literature.

## I. INTRODUCTION

While it is well known that finite-delay coding schemes do not achieve the asymptotic bounds in general (see, e.g., [1, Theorem 21] or [2]), the problem of obtaining the optimal coding schemes for finite delay is an important open problem with considerable practical implications [3]–[9]. Recently, there has been growing interest in utilizing zero-delay mappings in network applications, see, e.g., [10], [11] for coding over multiple access channels, [12]–[14] for distributed coding of correlated sources and [15], [16] for analog multiple description coding.

Until recently, there have been two main approaches to numerical optimization of the mappings: i) Optimization of the parameter set of a structured mapping [8], [9], [17], [18]. The performance of this approach is limited to the parametric form (structure) assumed. For example, in [19] saw-tooth like structure is assumed for the mapping in the Wyner-Ziv setting and parameters of such mapping are optimized. ii) Design based on power constrained channel optimized vector quantization where a discretized version of the problem is tackled using tools developed for vector quantization [5], [20], [21].

Our approach builds on the recent prior work in our lab [22] where the problem is studied in the original analog (functional) domain, i.e., without any discretization in the problem formulation and without any assumption of a parametrized mapping. In [22], necessary conditions for optimality of mappings were derived, noting that while such conditions have theoretical value, they generally identify local optima. They are practically useless in the case of highly complex cost surfaces, in other words, simple greedy methods that are based on iterative imposition of necessary conditions of optimality tend to get trapped in local minima. In [22], "noisy channel relaxation" (NCR) [23] was employed to mitigate this problem inherent to such optimization problems. As we show in this work, while NCR is rather sufficient for simple settings, using more advanced non-convex optimization tools improve the performance significantly in sophisticated network scenarios.

In this paper, we propose a method based on a powerful non-convex optimization framework, *deterministic annealing*, to numerically approach globally optimal zero-delay mappings in network scenarios. Our preliminary results appeared in [24], [25]. We particularly focus on scenarios given in Figure 1: The first case is a point-to-point source-channel coding with decoder side information (i.e., the decoder has access to side information that is correlated with the source). The second setting involves distributed (separate) coding and transmission of two correlated sources to a central decoder that reconstructs individual sources. We also consider the function computation problem, where the decoder estimates a function of the sources. This is of interest for certain applications such as a wireless sensor network deployed in order to compute a function of the measurements [26]–[30].

Deterministic annealing (DA) is derived within a probabilistic framework where the main idea is to introduce controlled randomization into the optimization process, yet deterministically optimize the appropriate expectation functionals. The application-specific cost is minimized at successive stages of decreasing randomness and nonrandom solution is obtained while avoiding many poor local minima. Based on information theoretic principles with analogies to statistical physics, DA has been successfully used in non-convex optimization problems including clustering [31], vector quantization [32], regression [33] and more (see review in [34]). We note that DA has been traditionally used in discrete settings such as quantizer optimization and integrating DA within the analog

Mustafa S. Mehmetoglu and Kenneth Rose are with the Department of Electrical and Computer Engineering, University of California, Santa Barbara, CA, 93106 USA, e-mail: {mehmetoglu, rose} @ece.ucsb.edu. Emrah Akyol is with the Coordinated Science Laboratory, University of Illinois at Urbana-Champaign, Urbana, IL, 61801, USA, e-mail: akyol@illinois.edu

This work was supported by the National Science Foundation under the grants CCF-1016861, CCF-1118075 and CCF-1320599. The material in this paper was presented in part at the IEEE Information Theory Workshop, Sevilla, Spain, September 2013 and the IEEE International Conference on Acoustics, Speech and Signal Processing, Florence, Italy, May 2014.

framework in here poses a significant challenge. There are many important advantages of the DA-based proposed method compared to prior work, including ability to avoid poor local minima and independence from initialization; and optimization in the original (analog) domain without any discretization or simplifying assumptions. Our approach improves significantly over prior approaches, some of which are NCR based [21], [22].

Having a powerful optimization method at hand, we analyze the structure of experimentally obtained mappings and investigate some conjectures made in prior work. For instance, one such claim was on the structure of optimal mappings in the side information setting, for which our results provide contradictory experimental evidence. Several practically important observations are made regarding the functional properties of the optimal mappings in network settings (see [35] for formal discussions of such properties in the point-to-point setting).

The rest of this paper is organized as follows. In Section II, we present preliminaries and the problem definition. In Section III and IV, we describe the proposed method. Experimental results are presented in Section V and concluding remarks are in Section VI.

## II. Preliminaries and Problem Definitions

### A. Notations

Let $\mathbb{R}$, $\mathbb{N}$, and $\mathbb{R}^+$ denote the respective sets of real numbers, natural numbers, and positive real numbers. We represent scalars and random variables with lowercase and uppercase letters (e.g., $x$ and $X$), column vectors and random column vectors with boldface lowercase and uppercase letters (e.g., $\boldsymbol{x}$ and $\boldsymbol{X}$), respectively. $\|\cdot\|$ denotes $L_2$ norm operator. Let $\mathbb{E}(\cdot)$ and $\mathbb{P}(\cdot)$ denote the expectation and probability operators, respectively. The probability density function of the random variable $X$ is $f_X(x)$. Let $\nabla$ and $\nabla_x$ denote the gradient and partial gradient with respect to $x$, respectively. Let $f'(x) = \frac{df(x)}{dx}$ denote the first-order derivative of the continuously differentiable function $f$. The Gaussian density with mean $\boldsymbol{\mu}$ and covariance matrix $R$ is denoted as $\mathcal{N}(\boldsymbol{\mu}, R)$. We use natural logarithms which, in general, may be complex, and the integrals are, in general, Lebesgue integrals.

### B. Problem Definition: Side Information

In the side information setting, given in Figure 1a, side information $\boldsymbol{Z} \in \mathbb{R}^{m_2}$ is available to the decoder, while source $\boldsymbol{X} \in \mathbb{R}^{m_1}$ is mapped to a channel input by the encoding function $\boldsymbol{g} : \mathbb{R}^{m_1} \to \mathbb{R}^p$ and transmitted over the channel with additive noise $\boldsymbol{N} \in \mathbb{R}^p$. The received channel output $\boldsymbol{Y} = \boldsymbol{g}(\boldsymbol{X}) + \boldsymbol{N}$ and side information $\boldsymbol{Z}$ are mapped to the estimate $\hat{\boldsymbol{X}}$ by the decoding function $\boldsymbol{w} : \mathbb{R}^p \times \mathbb{R}^{m_2} \to \mathbb{R}^{m_1}$. The problem is to find optimal mappings $\boldsymbol{g}, \boldsymbol{w}$, where optimality is in the sense that they minimize MSE

$$D(\boldsymbol{g}, \boldsymbol{w}) = \mathbb{E}\{\|\boldsymbol{X} - \hat{\boldsymbol{X}})\|^2\}, \quad (1)$$

subject to some power constraint on the encoder

$$P(\boldsymbol{g}) = \mathbb{E}\{\|\boldsymbol{g}(\boldsymbol{X})\|^2\} \leq P_E \quad (2)$$

where $P_E > 0$ is the specified encoder power level. Simple time-sharing arguments show that $D$ is a convex functional of $P$, hence the solution is achieved at $P = P_E$ (see [35] for details.) Converting to Lagrangian formulation, we define the following cost to be minimized

$$J = D(\boldsymbol{g}, \boldsymbol{w}) + \lambda(P(\boldsymbol{g}) - P_E) \quad (3)$$

where $\lambda$ is a Lagrange multiplier corresponding to the power constraint on the encoder (we suppressed the dependence of $J$ on $\boldsymbol{g}$ and $\boldsymbol{w}$).

### C. Problem Definition: Distributed Coding

The distributed coding setting, given in Figure 1b, has two sources $\boldsymbol{X}_1 \in \mathbb{R}^{m_1}$ and $\boldsymbol{X}_2 \in \mathbb{R}^{m_2}$ mapped to some channel input by the encoding functions $\boldsymbol{g}_i : \mathbb{R}^{m_i} \to \mathbb{R}^{p_i}$, and the decoder receives $\boldsymbol{Y}_i = \boldsymbol{g}_i(\boldsymbol{X}_i) + \boldsymbol{N}_i$ for $i = 1, 2$. In general, the decoder might have two type of objectives. In the first one, the decoder aims to reconstruct each source with minimum distortion. The decoder is defined as $\boldsymbol{w} : \mathbb{R}^{p_1} \times \mathbb{R}^{p_2} \to \mathbb{R}^{m_1} \times \mathbb{R}^{m_2}$ as it maps the received channel outputs to the estimates $\hat{\boldsymbol{X}}_i$ for $i = 1, 2$. For this case, we define distortion as

$$D(\boldsymbol{g}_1, \boldsymbol{g}_2, \boldsymbol{w}) = \mathbb{E}\{\|\boldsymbol{X}_1 - \hat{\boldsymbol{X}}_1\|^2 + \eta\|\boldsymbol{X}_2 - \hat{\boldsymbol{X}}_2\|^2\} \quad (4)$$

where $\eta \in \mathbb{R}^+$ is a given weight coefficient. The second type of problems involve function computation. Denoting the desired function as $\boldsymbol{\gamma}(\boldsymbol{X}_1, \boldsymbol{X}_2) : \mathbb{R}^{m_1} \times \mathbb{R}^{m_2} \to \mathbb{R}^r$, the decoder is defined as $\boldsymbol{w} : \mathbb{R}^{p_1} \times \mathbb{R}^{p_2} \to \mathbb{R}^r$ and the cost is given by

$$D(\boldsymbol{g}_1, \boldsymbol{g}_2, \boldsymbol{w}) = \mathbb{E}\{\|\boldsymbol{\gamma}(\boldsymbol{X}_1, \boldsymbol{X}_2) - \boldsymbol{w}(\boldsymbol{Y}_1, \boldsymbol{Y}_2)\|^2\}. \quad (5)$$

The problem, for both cases, is to find the mappings $\boldsymbol{g}_1, \boldsymbol{g}_2, \boldsymbol{w}$ that minimize the overall distortion (which is given in (4) or (5) depending on the objective) subject to power constraints on the encoders, which can be in two forms: Individual power constraints given by

$$P(\boldsymbol{g}_i) = \mathbb{E}\{\|\boldsymbol{g}_i(\boldsymbol{X}_i)\|^2\} \leq P_{T,i} \text{ for } i = 1, 2. \quad (6)$$

or a total power constraint of the form

$$\sum_{i=1}^{2} P(\boldsymbol{g}_i) \leq P_T, \quad (7)$$

which offers the additional degree of freedom of optimizing power allocations to the encoders. For optimization purposes, we similarly define the following Lagrangian functional as the objective cost to be minimized

$$J = D + \sum_{i=1}^{2} \lambda_i (P(\boldsymbol{g}_i) - P_{T,i}), \quad (8)$$

where $\lambda_i \in \mathbb{R}^+$, $i = 1, 2$, are Lagrange multipliers to impose the individual power constraints on the encoders in the first case. The total power constraint case corresponds to the special case of (8) with $\lambda_1 = \lambda_2 = \lambda$, i.e., the Lagrangian cost to minimize is

$$J = D + \lambda(P(\boldsymbol{g}_1) + P(\boldsymbol{g}_2) - P_T), \quad (9)$$

where $\lambda$ controls the total power.

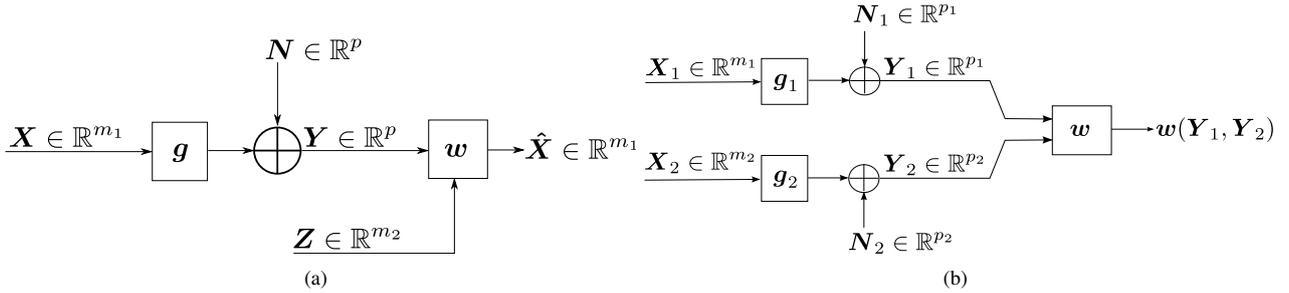

Fig. 1. Problem settings that we consider. (a) Decoder with side information. (b) Distributed coding setting.

## D. Prior Work: Necessary Conditions of Optimality and Greedy Descent Algorithms

Here, we summarize the relevant contributions of prior work (see [22] for more details). For the side information setting, let the encoder $g$ be fixed. Then, the optimal decoder is the MSE estimator of $X$ given $Z = z$ and $Y = y$:

$$w(y, z) = \mathbb{E}\{X|y, z\}. \quad (10)$$

Expanding the expressions for expectation and applying Bayes' rule, the optimal decoder can be written in terms of known quantities as

$$w(y, z) = \frac{\int x\, f_{X,Z}(x,z)\, f_N(y - g(x))\, \mathrm{d}x}{\int f_{X,Z}(x,z)\, f_N(y - g(x))\, \mathrm{d}x}, \quad (11)$$

where we used the fact that $f_{Y|X}(y, x) = f_N(y - g(x))$. For optimality of $g$, assuming the decoder $w$ is fixed, a necessary condition is

$$\nabla_g J(g, w) = 0, \quad (12)$$

where

$$\nabla_g J(g,w) = \lambda f_X(x) g(x) \\ - \mathbb{E}\{w'(g(x)+N,Z)(x-w(g(x)+N,Z))\}, \quad (13)$$

and $w'$ denotes the Jacobian of $w$ with respect to its first argument (see [22] for proof).

**Remark 1.** *Note that in the case of jointly Gaussian sources and Gaussian channel(s) with matched source-channel dimensions, linear mappings satisfy the necessary conditions of optimality, however, they are highly suboptimal, see, e.g., [22]. As we will see, careful optimization obtains considerably better mappings that are far from linear.*

The necessary conditions of optimality for the distributed coding setting can be derived similarly, and are omitted for brevity, see [22]. Iteratively alternating between the imposition of individual necessary conditions of optimality will successively decrease the Lagrangian cost until a stationary point is reached. We refer to this method as "greedy descent". There is no reason to expect that a greedy descent algorithm will converge to the globally optimal solution. In fact, experiments show severe issues of local optima and strong dependence on initialization of such methods. As a remedy, the noisy channel relaxation (NCR) method of [23] was embedded in the algorithm in [22], i.e., the descent method was run at gradually decreasing levels of $\lambda$, wherein the result at each level serves as initialization for the next level of $\lambda$ (see [23] for details). While such simple relaxations are effective in simple communication settings, the networked problems we consider here require a stronger optimization approach.

## E. Asymptotically Achievable Limits

It is insightful to consider asymptotic bounds, which are obtained at infinite delay, while keeping in mind that the problem we consider is delay limited. Let $R(D)$ and $C(P)$ denote the source rate-distortion function and channel capacity, respectively. According to Shannon's source and channel coding theorems, the source can be compressed to $R(D)$ bits (per source sample) at distortion level $D$, and that $C(P)$ bits can be transmitted over the channel (per channel use) with arbitrarily low probability of error (see, e.g., [36]). The optimal coding scheme is the tandem combination of the optimal source and channel coding schemes, hence, by setting

$$R(D) = C(P), \quad (14)$$

one obtains a lower bound on the distortion of any source-channel coding scheme. For simplicity, we derive the expressions for the "optimum performance theoretically attainable" (OPTA) for Gaussian scalar source and noise. The channel capacity with additive white Gaussian noise is given by

$$C(P) = \frac{1}{2}\log\left(1 + \frac{P}{\sigma_N^2}\right), \quad (15)$$

where $P$ is the transmission power and $\sigma_N^2$ is the noise variance.

For source-channel coding with decoder side information, OPTA can be obtained by equating Wyner-Ziv rate distortion function [37] to the channel capacity. The Wyner-Ziv rate distortion function of $X$, when $Z$ serves as side information, and $(X, Z) \sim \mathcal{N}(\mathbf{0}, R_{X,Z})$ where $R_{X,Z} = \sigma_X^2 \begin{bmatrix} 1 & \rho \\ \rho & 1 \end{bmatrix}$ and $\sigma_X^2$, $\rho$ are the variance and correlation coefficient, respectively, with $|\rho| \leq 1$ is:

$$R(D) = \max\left(0, \frac{1}{2}\log\frac{(1-\rho^2)\sigma_X^2}{D}\right). \quad (16)$$

We plug (16) and (15) in (14) to obtain

$$D_{OPTA} = \frac{(1-\rho^2)\sigma_X^2}{\left(1 + \frac{P_T}{\sigma_N^2}\right)}. \quad (17)$$

For quadratic Gaussian distributed source coding for sources $(X_1, X_2) \sim \mathcal{N}(\mathbf{0}, R_{X_1, X_2})$ where $R_{X_1, X_2} = \sigma_X^2 \begin{bmatrix} 1 & \rho \\ \rho & 1 \end{bmatrix}$ with $|\rho| \leq 1$, the complete rate distortion region satisfies the following inequalities [38]:

$$R_1 \geq \frac{1}{2} \log^+ \left( \frac{1 - \rho^2 + \rho^2 2^{-2R_2}}{D_1} \right) \tag{18}$$

$$R_2 \geq \frac{1}{2} \log^+ \left( \frac{1 - \rho^2 + \rho^2 2^{-2R_1}}{D_2} \right) \tag{19}$$

$$R_1 + R_2 \geq \frac{1}{2} \log^+ \left( \frac{(1 - \rho^2) \beta(D_1, D_2)}{2 D_1 D_2} \right) \tag{20}$$

where $\log^+ x = \max(0, \log x)$ and

$$\beta(D_1, D_2) = 1 + \sqrt{1 + \frac{4\rho^2 D_1 D_2}{(1 - \rho^2)^2}}. \tag{21}$$

We set $R_i = C(P_i)$ for $i = 1, 2$, where $C(P)$ is given in (15) to obtain OPTA.

## III. PROPOSED METHOD FOR SIDE INFORMATION SETTING

### A. Overview

In this section, we develop the DA based method for the optimization of encoder and decoder mappings. Since the decoder is given in closed form, the method focuses on optimizing the encoder mapping. We first partition the input space of the encoder into partition cells and assign a local model to each of the cells. Next, the encoder output is made probabilistic by randomizing the partitions, i.e., input points are assigned to each local model according to some probability distribution. We then propose an optimization process where the (random) encoder is optimized (along with the decoder) while constraining the Shannon entropy. By gradually reducing the entropy to 0, we obtain the desired mappings.

### B. Derivation of proposed method

We consider piecewise functions which approximate the desired mappings by partitioning the space and matching a simple local model to each region. Piecewise functions consist of two components: a space partition and a parametric local model per partition cell. First, the source space $\mathbb{R}^m$ is partitioned into $\mathcal{K}$ regions (cells) denoted $\mathcal{R}_k^m$. Each cell $\mathcal{R}_k^m$ has an associated function $\boldsymbol{g}_k$ which is parametrized (affine, lattice, etc.) and the parameter set is denoted by $\Lambda_k$. Thus, the encoding function can be written as

$$\boldsymbol{g}(\boldsymbol{x}) = \boldsymbol{g}_k(\boldsymbol{x}) \text{ for } \boldsymbol{x} \in \mathcal{R}_k^m \text{ and for } k = 1, \ldots, \mathcal{K} \tag{22}$$

In (22), the selection of local model index $k$ is deterministic for a given realization of $\boldsymbol{X}$, i.e., the output of the encoder only depends on $\boldsymbol{X}$. To derive a DA based approach, we introduce a random variable, $K$, that corresponds to random selection of index $k$. In other words, let the encoder randomly select the local model index $k$ when it receives an input $\boldsymbol{x}$, according to the value of a random variable that we call $K$. For a given realization of $\boldsymbol{X}$, the output of the encoder is now given in probability as

$$\boldsymbol{g}(\boldsymbol{x}) = \boldsymbol{g}_k(\boldsymbol{x}) \text{ with probability } p_{K|\boldsymbol{X}}(k|\boldsymbol{x}). \tag{23}$$

The conditional probability $p_{K|\boldsymbol{X}}(k|\boldsymbol{x})$ is referred to as *association probability*, in the sense that it represents the probability of input point $\boldsymbol{x}$ belonging to cell $\mathcal{R}_k^m$ (thus, the source space partition is now random). The probability distribution that we introduce (and optimize) is $p_{K|\boldsymbol{X}}$ (not the joint $p_{\boldsymbol{X}, K}$) since the input distribution is given in the problem statement and is therefore fixed. The MSE cost and transmission power are still calculated as in (1) and (2), though the expectation is now taken over $K$ in addition to what was done before. Let us now fix $\Lambda_k$ and $\boldsymbol{w}$, and consider optimizing (3) with respect to $p_{K|\boldsymbol{X}}$. It is clear that the optimal $p_{K|\boldsymbol{X}}$ will implement 'hard' associations, that is, every point $\boldsymbol{x}$ will be fully associated with the local model that makes the minimum contribution to cost[1]. Although this is desirable eventually, in order to avoid poor local optima we impose and control the level of randomness, i.e., we introduce a constraint on the randomness of the encoder, which is measured by the Shannon entropy. The total entropy of the encoder is given by $H(\boldsymbol{X}, K) = H(\boldsymbol{X}) + H(K|\boldsymbol{X})$ and since $H(\boldsymbol{X})$ is constant (predetermined by the source), the entropic quantity of interest is the conditional entropy $H(K|\boldsymbol{X})$. This is also intuitively justified in the sense that the randomness we introduced into the problem is precisely captured by $p_{K|\boldsymbol{X}}$, hence can be measured and controlled by $H(K|\boldsymbol{X})$. We denote the randomness of the solution by $H$ and define it as $H \triangleq H(K|\boldsymbol{X})$ where

$$H(K|\boldsymbol{X}) = -\mathbb{E}\{\log p_{K|\boldsymbol{X}}\}. \tag{24}$$

The problem is now recast as minimization of the expected cost with respect to parameters of local models, association probabilities and decoder, subject to a constraint on the level of randomness of the system, i.e.,

$$\underset{\Lambda_1, \ldots, \Lambda_{\mathcal{K}}, p(1|\boldsymbol{x}), \ldots, p(\mathcal{K}|\boldsymbol{x}), \boldsymbol{w}}{\text{minimize}} \quad J,$$
$$\text{subject to} \quad H \geq H_0,$$

where $J$ is defined in (3) and $H_0$ specifies the minimum requirement on the entropy level. This constrained optimization problem can be reformulated by introducing Lagrange parameter $T \in \mathbb{R}^+$ to obtain the Lagrangian

$$F = J - T(H - H_0), \tag{25}$$

to be minimized. There are two important extremal points of this Lagrangian. First, for $T \to \infty$, the minimum $F$ is obtained by maximizing the entropy, which is achieved by uniform association probabilities: $p_{K|\boldsymbol{X}}(k|\boldsymbol{x}) = 1/\mathcal{K}$ for all $k$ and $\boldsymbol{x}$. Consequently, all local models equally account for all points and are identical once optimized, or effectively, there is a single *distinct* local model. Secondly, in the limit $T \to 0$, minimizing $F$ corresponds to minimizing $J$ directly, which produces a deterministic encoder. This intuitive observation can be verified by the expression for optimal $p_{K|\boldsymbol{X}}(k|\boldsymbol{x})$ given

---
[1]Therefore, the generalized search space of random encoders have the same global minimum as the original problem.

in Section III-D.

Although DA is derived from information theoretic principles, it is motivated by and has strong analogies to annealing processes in statistical physics (see [34] for details). We accordingly refer to the Lagrangian functional in (25) as (Helmholtz) free energy, and Lagrange parameter $T$ as "temperature".

*C. Deterministic Annealing*

The optimization method starts at a high value of $T$ and gradually lowers it while minimizing $F$ at each step. At high temperature, there is effectively a single distinct local model. As the temperature is decreased, a bifurcation point is reached where the current solution is no longer a minimum, so that there exists a better solution with a higher number of distinct local models. Intuitively, at this temperature, the current solution is a saddle point where multiple local models are coincident (i.e., their parameters are same) and in order to move to a better solution, it is necessary to perturb the local models. Such bifurcations are referred to as "phase transitions" and the corresponding temperatures are called "critical temperatures"[2].

We present an example simulation in Figure 2 that illustrates the basics of the method, including phase transitions. Here the sources and channel are scalar, i.e., $m = n = 1$, $g_k$ are selected as affine and $\mathcal{K} = 2$. When $T$ is large, there is a single distinct local model. As we lower $T$, the system goes through a phase transition where the two local models split from each other (after a slight perturbation). The corresponding value of $T$ is referred to as the first critical temperature. Note how entropy ($H$) is traded for reduction in cost ($J$).

Mappings with more than 2 local models can be obtained by starting with a larger $\mathcal{K}$. However, a computationally more efficient method that we employ here is as follows: We start with 1 local model and keep only the distinct local models, but duplicate and perturb them at each temperature. The duplicates will merge at every iteration until a critical temperature is reached, and will split into distinct models at a phase transition.

Although our method is derived in the general, continuous source and channel domain, in practical simulations we sample the source and noise distributions to allow numerical computation of integrals. The sampling is not "inherent" to the derived method and, in fact, can be adjusted during the algorithm run. We emphasize that this is in contrast with prior quantizer design based methods that are entirely formulated in a discrete setting.

The practical algorithm is initialized with a single local model. Since $T$ must be set higher than the first critical temperature, we simply choose $T$ large enough that during the first couple of temperatures, duplicated local models merge back, i.e., no phase transitions are observed. As the temperature is gradually lowered, we track the minimum, i.e., find the association probabilities $p_{K|X}(k|\boldsymbol{x})$, local model parameters $\Lambda_k$ and decoder $\boldsymbol{w}$ that minimize the Lagrangian $F$.

[2] We omit the derivation of critical temperatures in this paper, see [34] for phase transition analysis in the general DA setting.

As demonstrated, the system will go through phase transitions during which the number of local models, $\mathcal{K}$, increases. We stop when $T$ is near 0 and perform "zero entropy iteration", i.e., associate every source point with the "best" local model to obtain deterministic encoder. We accordingly give a brief sketch of the practical method in Algorithm 1. In Step 6, we employed an exponential cooling schedule. Update equations for Step 3 are given in the next section.

---

**Algorithm 1** Proposed DA-Based Method

**Inputs:** *Involved distributions, desired local model type, $\lambda$, $\alpha$, $\epsilon$, $\Delta_F$, $T_{min}$, $\Delta_{\boldsymbol{g}}$.*
**Outputs:** *Optimized $\boldsymbol{g}, \boldsymbol{w}$.*
**Initialization:** $T = T_{max}$, $\mathcal{K} = 1$, *randomly chosen $\boldsymbol{g}_1$. $J_{old} = J_{initial}$.*

1. **Duplication:**
   For each $\boldsymbol{g}_i$, create an identical local model $\boldsymbol{g}_j$.
   $p(i|\boldsymbol{x}) \leftarrow \frac{p(i|\boldsymbol{x})}{2}$ and $p(j|\boldsymbol{x}) \leftarrow \frac{p(i|\boldsymbol{x})}{2}$.
   $\mathcal{K} \leftarrow 2\mathcal{K}$.

2. **Perturbation:**
   For each parameter $\phi_k \in \Lambda_k$, $\phi_k \leftarrow \phi_k + \epsilon R$, where $R$ is standard Gaussian random variable.

3. **Thermal Equilibrium:**
   Compute $F$ and set $F_{old} \leftarrow F$.
   3.1. Compute optimal $\boldsymbol{w}$ using (30).
   3.2. Compute optimal $p(k|\boldsymbol{x})$, $\forall k$ using (26).
   3.3. Optimize $\Lambda_k$, $\forall k$ using (28).
   3.4. Compute $F$. If $\frac{F - F_{old}}{F_{old}} \leq \Delta_F$, go to Step 4, otherwise $F_{old} \leftarrow F$ and go to Step 3.1.

4. **Model Size:**
   If $d(\Lambda_i, \Lambda_j) < \Delta_{\boldsymbol{g}}$, where $d(\cdot, \cdot)$ is euclidean distance, remove $\boldsymbol{g}_j$ and set $p(i|\boldsymbol{x}) \leftarrow p(i|\boldsymbol{x}) + p(j|\boldsymbol{x})$, $\forall i, j$.
   $\mathcal{K} \leftarrow$ New model size.

5. **Stopping:**
   Stop if $T \leq T_{min}$, otherwise go to Step 6.

6. **Cooling:**
   $T \leftarrow T * \alpha$.
   Go to Step 1.

---

*D. Update Equations*

The central part of the method is the minimization of free energy ($F$) by iteratively updating the association probabilities, local model parameters and decoders. The following theorem, whose proof is presented in the Appendix, states the update equations for association probabilities.

**Theorem 1.** *At any temperature $T$, minimum free energy $F$ is achieved when association probabilities are in the form of Gibbs distribution given as:*

$$p(k|\boldsymbol{x}) = \frac{e^{-J_k(\boldsymbol{x})/T}}{\sum_{k'} e^{-J_{k'}(\boldsymbol{x})/T}} \quad \forall k, \quad (26)$$

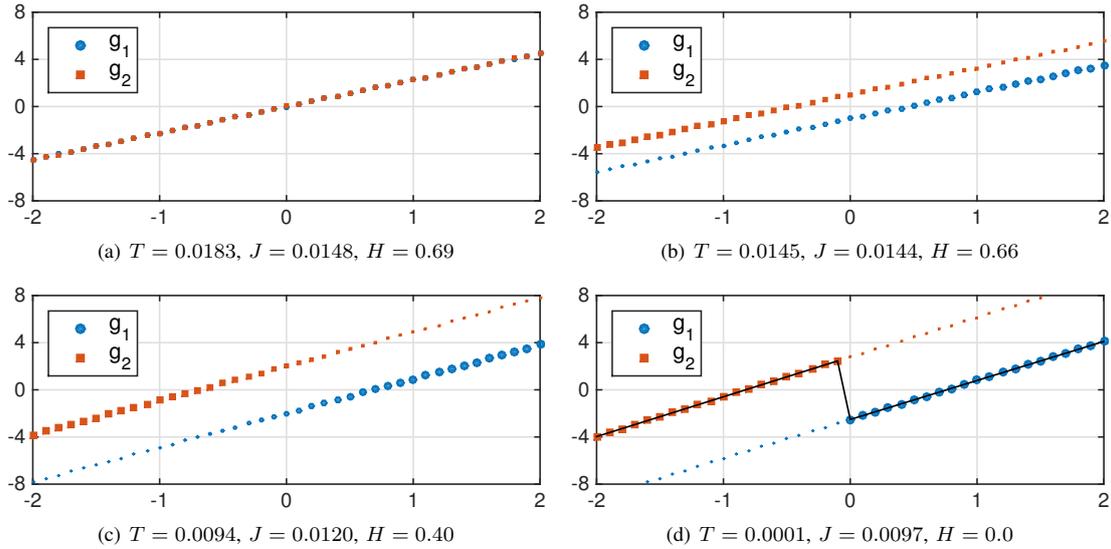

Fig. 2. The evolution of the encoder in the algorithm is demonstrated. The two models are shown by dotted lines and the sizes of dots are relative to the probability association at that input point. The line in (d) is the deterministic encoder obtained. $\mathcal{K} = 2$.

where $J_k(\boldsymbol{x})$ is given by

$$J_k(\boldsymbol{x}) = \mathbb{E}\{\|\boldsymbol{x} - \boldsymbol{w}(\boldsymbol{g}_k(\boldsymbol{x}) + \boldsymbol{N}, \boldsymbol{Z})\|^2\} + \lambda \|\boldsymbol{g}_k(\boldsymbol{x})\|^2. \quad (27)$$

**Remark 2.** *Theorem 1 is analogous to the principle of minimal free energy in statistical physics. A fundamental principle in statistical physics states that the minimum free energy is achieved when the system is at thermal equilibrium, at which point it is governed by Gibbs distribution.*

The evolution of association probabilities, $p(k|\boldsymbol{x})$, during the annealing process can be observed from how (26) is changing with $T$. The following corollary confirms the intuitive explanation we provided earlier.

**Corollary 1.** *As $T \to \infty$ (at a high temperature) the system is governed by uniform association probabilities and the entropy is maximum. As $T \to 0$, the associations become deterministic and the entropy is $0$.*

The optimal local model parameters cannot be obtained in closed form, hence we perform gradient descent search. A local model parameter $\phi_k \in \Lambda_k$ is updated according to

$$\phi_k \leftarrow \phi_k - \varphi \frac{\partial F}{\partial \phi_k} \quad (28)$$

where $\varphi$ is selected by line search and the gradient can be obtained as

$$\frac{\partial F}{\partial \phi_k} = \frac{\partial J}{\partial \phi_k} = \int_{\boldsymbol{x}} f_{\boldsymbol{X}}(\boldsymbol{x}) p(k|\boldsymbol{x}) \frac{\partial J_k(\boldsymbol{x})}{\partial \phi_k} \, d\boldsymbol{x}. \quad (29)$$

The derivative $\frac{\partial J_k(\boldsymbol{x})}{\partial \phi_k}$ is calculated numerically. The optimal decoder can be derived similar to (11):

$$\boldsymbol{w}(\boldsymbol{y}, \boldsymbol{z}) = \frac{\int \boldsymbol{x} f_{\boldsymbol{X}, \boldsymbol{Z}}(\boldsymbol{x}, \boldsymbol{z}) \sum_k f_{\boldsymbol{N}}(\boldsymbol{y} - \boldsymbol{g}_k(\boldsymbol{x})) p(k|\boldsymbol{x}) \, d\boldsymbol{x}}{\int f_{\boldsymbol{X}, \boldsymbol{Z}}(\boldsymbol{x}, \boldsymbol{z}) \sum_k f_{\boldsymbol{N}}(\boldsymbol{y} - \boldsymbol{g}_k(\boldsymbol{x})) p(k|\boldsymbol{x}) \, d\boldsymbol{x}}. \quad (30)$$

*E. Design Complexity*

Due to difficulties in estimating the time required for gradient descent, exact comparison of computational complexity of numerical optimization methods (including the method presented here and others referred to in Section II-D) is difficult and depends on the actual source-channel distributions as well as choice of various algorithm parameters. On the other hand, optimization of parametrized mappings (e.g., in [13]) is faster, but requires knowing the structure of a good solution, which can be obtained by methods such as the one presented here. In our experiments, the time required for DA was on the same order as that of NCR, albeit with a higher constant. Thus, better performance is obtained at the expense of slight increase in complexity.

## IV. METHOD FOR DISTRIBUTED CODING

Although the method described in the previous section can be used for optimizing the distributed encoders separately (within separate annealing processes), we found that such a method fails to avoid poor local minima as it fails to account for interaction between encoder optimizations. Instead, we develop a method here that optimizes the (random) encoders and decoders within a single annealing process. The resulting annealing method is a direct extension of the previous one, albeit with higher complexity due to the distributed nature of the problem.

We have two independent sets of partitions of input source space: $\mathcal{K}_1$ cells represented by $\mathcal{R}^m_{k_1}$ and $\mathcal{K}_2$ cells represented by $\mathcal{R}^m_{k_2}$. We define both encoders in this setting as

$$\boldsymbol{g}_i(\boldsymbol{x}_i) = \boldsymbol{g}_{k_i}(\boldsymbol{x}_i) \text{ for } \boldsymbol{x}_i \in \mathbb{R}^m_{k_i}, \; i = 1, 2. \quad (31)$$

Following the same procedure of randomization, we define random variables $K_1$ and $K_2$ along with association probabilities:

$$p(k_i|\boldsymbol{x}_i) \triangleq \mathbb{P}\{\boldsymbol{x}_i \in \mathcal{R}^m_{k_i}\}, \quad \forall k_i, \boldsymbol{x}_i, \text{ for } i = 1, 2. \quad (32)$$

The cost is to be minimized subject to the constraint on the joint entropy of the system. Noting that $K_1 \leftrightarrow X_1 \leftrightarrow X_2 \leftrightarrow K_2$ form a Markov chain by construction, we express the joint entropy as

$$H(\boldsymbol{X}_1, K_1, \boldsymbol{X}_2, K_2) = H(\boldsymbol{X}_1, \boldsymbol{X}_2) + H(K_1|\boldsymbol{X}_1) + H(K_2|\boldsymbol{X}_2). \quad (33)$$

Since $H(\boldsymbol{X}_1, \boldsymbol{X}_2)$ is a constant determined by the sources, we define $H \triangleq H(K_1|\boldsymbol{X}_1) + H(K_2|\boldsymbol{X}_2)$ where

$$H(K_i|\boldsymbol{X}_i) = \mathbb{E}\{\log p(K_i|\boldsymbol{X}_i)\} \text{ for } i = 1, 2, \quad (34)$$

and the free energy of the system is given by (25).

The algorithm sketch is similar to the side information setting and is not reproduced here. Since we optimize both encoders within the same annealing process, the same operations in the Algorithm are performed for both encoders, sequentially. The following theorem presents the optimal association probabilities for the distributed setting. The proof follows from the steps in the proof of Theorem 1 and omitted for brevity.

**Theorem 2.** *At any temperature, minimum free energy (F) is achieved when the system is governed by Gibbs distribution given as:*

$$\mathrm{p}(k_i|\boldsymbol{x}_i) = \frac{e^{-J_{k_i}(\boldsymbol{x}_i)/T}}{\sum_{k'_i} e^{-J_{k'_i}(\boldsymbol{x}_i)/T}} \quad \text{for } i = 1, 2 \quad (35)$$

*where*

$$J_{k_i}(\boldsymbol{x}_i) = \mathbb{E}\{\|\boldsymbol{X}_1 - \hat{\boldsymbol{X}}_1\|^2 + \eta\|\boldsymbol{X}_2 - \hat{\boldsymbol{X}}_2\|^2 | \boldsymbol{X}_i = \boldsymbol{x}_i, K_i = k_i\}$$
$$+ \lambda_i \boldsymbol{g}_{k_i}^2(\boldsymbol{x}_i) \quad (36)$$

*if the cost is defined as in (4), and*

$$J_{k_i}(\boldsymbol{x}_i) = \mathbb{E}\{\|\gamma(\boldsymbol{X}_1, \boldsymbol{X}_2) - \boldsymbol{w}(\boldsymbol{Y}_1, \boldsymbol{Y}_2)\|^2 | \boldsymbol{X}_i = \boldsymbol{x}_i, K_i = k_i\}$$
$$+ \lambda_i \boldsymbol{g}_{k_i}^2(\boldsymbol{x}_i) \quad (37)$$

*if the cost is defined as in (5).*

The parameters of local models can be optimized through gradient descent search. Optimal decoding is achieved similarly as $\hat{\boldsymbol{X}}_i = \mathbb{E}\{\boldsymbol{X}_i|\boldsymbol{y}_1, \boldsymbol{y}_2\}$ for $i = 1, 2$ for first type of objective, and $\boldsymbol{w}(\boldsymbol{y}_1, \boldsymbol{y}_2) = \mathbb{E}\{\gamma(\boldsymbol{X}_1, \boldsymbol{X}_2)|\boldsymbol{y}_1, \boldsymbol{y}_2\}$ for the second type. Both expressions can be written in terms of known quantities similar to that in (11).

## V. EXPERIMENTAL RESULTS

While the proposed algorithm is general and directly applicable to any choice of source and channel dimensions, for conciseness of the results section, we assume that sources and channels are scalar. In this case, the encoder mapping is denoted as $g : \mathbb{R} \to \mathbb{R}$ and the local model functions $g_k$ are selected as affine. In principle, the set of $g_k$ can be chosen from any parametric model. Choosing a more complex model, such as a higher order polynomial, can potentially improve the performance of the algorithm, albeit with increased computational complexity. For the exponential cooling schedule, we set $\alpha = 0.95$, i.e., $T \leftarrow T*0.95$. The performance of the proposed method is assessed by comparisons to the optimal affine solution, greedy method and NCR-based method developed in

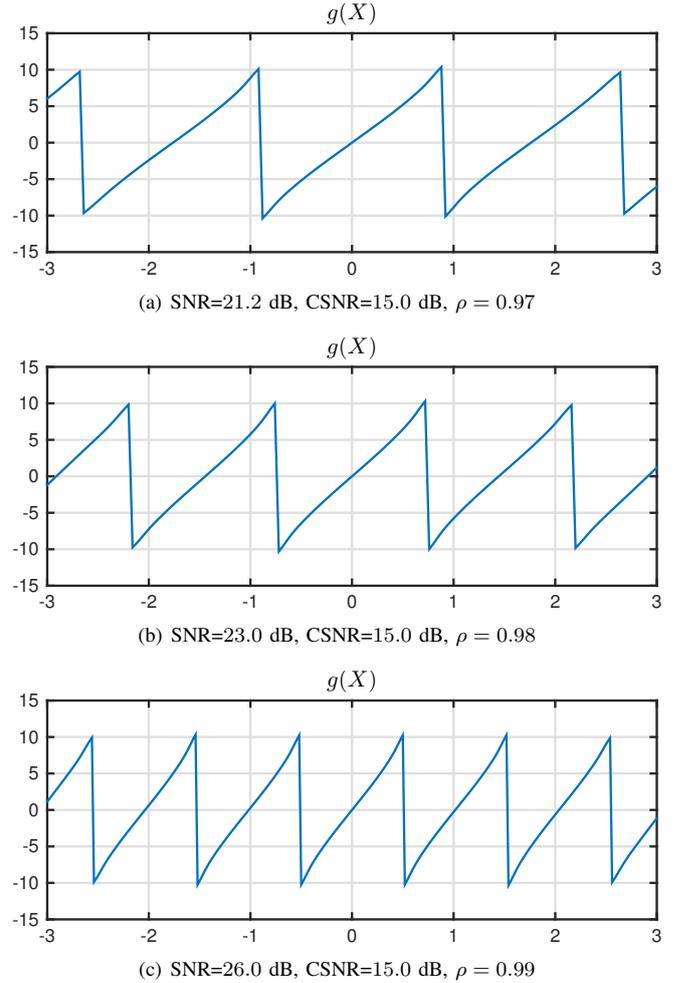

Fig. 3. Example encoder mappings, generated by DA, for the decoder side information setting, jointly Gaussian source and side information.

[22], as well as OPTA (for reference only, as OPTA requires infinite delay). For the NCR based method, we decrease $\lambda$ (in distributed coding, we decrease $\lambda_1$ and $\lambda_2$ simultaneously) exponentially as $\lambda_{new} = \lambda_{old} * 0.8$ in 50 steps to the desired value.

The noise signals in all examples are chosen as independent zero-mean Gaussians with unit variance, i.e., $N \sim \mathcal{N}(0, 1)$, $N_1 \sim \mathcal{N}(0, 1)$, $N_2 \sim \mathcal{N}(0, 1)$. For numerical computations, we sample the source and noise distributions on a uniform grid with spacing $\Delta = 0.02$. We also impose bounded support ($-5\sigma$ to $+5\sigma$), i.e., we neglect tails of infinite support distributions in the examples.

### A. Side Information Setting

We first give examples for the Gaussian case, where the source and side information are jointly Gaussian, distributed according to $\mathcal{N}(\boldsymbol{\mu}, R)$ where $\boldsymbol{\mu} = [0, 0]$, $R = \begin{bmatrix} 1 & \rho \\ \rho & 1 \end{bmatrix}$, and $|\rho| < 1$ is the correlation coefficient between source and side information. We define SNR $= 10 \log_{10}(1/D)$ and CSNR $= 10 \log_{10}(P(g))$.

Example mappings are given in Figure 3. We first note that the central characteristics observed in digital Wyner-Ziv mappings are captured by analog mappings as noted before (see, e.g., [21], [22]), in the sense of many-to-one mappings, where multiple source intervals are mapped to the same channel interval. We refer to each one-to-one section in these mappings as a "bin", in Figure 3a there are 5 bins in the interval shown (the meaning of bin here is different than in digital Wyner-Ziv mappings). The uncertainty about the source interval is resolved (significantly decreased) by the decoder using the side information. Since all variables are Gaussian and distortion measure is MSE, it is intuitively intriguing to investigate whether the optimal mappings have any parametric form or structure to be exploited in the design stage. For example, since in the absence of decoder side information optimal mappings are well known to be linear, one can expect to see linear mappings in each bin. In fact, such parametric form was explicitly assumed in [19], and it was reported the optimized parametric mappings perform very close to the results obtained via NCR in [22]. Our numerical results demonstrate that each bin is non-linear as some nonlinearity can be observed especially near the ends of each bin, as opposed to the conjecture in [22].

From Figure 3 we see how the width of bins depends on the correlation between the source and side information. It can be seen that at higher correlation the bins are narrower. This is intuitively expected since, as the correlation increases, so does the benefit of side information in terms of distinguishing different bins. To exploit this capability, the encoder narrows the bins, which in turn reduces the power $\mathbb{E}\{g^2(X_1)\}$.

To illustrate the improvement of DA over NCR in the encoding mappings themselves, we present two mappings obtained by NCR in Figure 4. We emphasize that the performance of NCR depends on initial mappings, initial noise level and the noise-relaxation schedule. This dependence is illustrated in Figure 4, where in one case the shape of bins are different then those in DA and sub-optimal, and in the other the points of discontinuity are not optimal.

We also give an example with a different source distribution, Gaussian mixture, in Figure 5:

$$(X_1, X_2) \sim \left(\frac{1}{2}\mathcal{N}(\boldsymbol{\mu}_1, R) + \frac{1}{2}\mathcal{N}(\boldsymbol{\mu}_2, R)\right) \quad (38)$$

where $\boldsymbol{\mu}_1 = [-3, -3]$, $\boldsymbol{\mu}_2 = [3, 3]$ and $R = \begin{bmatrix} 1 & 0.95 \\ 0.95 & 1 \end{bmatrix}$. This distribution has two Gaussian "nodes" centered far from each other at $x = -3$ and $x = 3$. From an intuitive point of view, the optimum encoder can be viewed as two Wyner-Ziv like encoders, occupying the negative and positive halves of real line and both centered at the node centers. It is clear that for several source and channel distributions, optimal encoding mappings are many-to-one, i.e., this property is not unique to the Gaussian distribution.

The comparative performance results for different optimization techniques is given in Figure 6 for correlation coefficient $\rho = 0.99$. Since NCR performance depends on the initial conditions, we ran the NCR algorithm several times with different conditions and pick the mappings with best performance. Results from the greedy method are also presented in order to illustrate the abundance of locally optimum points and the difficulty of the optimization problem. Note that the proposed method is independent of the initialization and only run once. We also present the performance of OPTA as benchmark while noting that it is asymptotic and may require infinite delay. The performance of linear encoder and decoder is plotted as well, since it is also a local minimum (see Remark 1). It is important to note that the linear solution performs significantly worse than the non-linear mappings obtained.

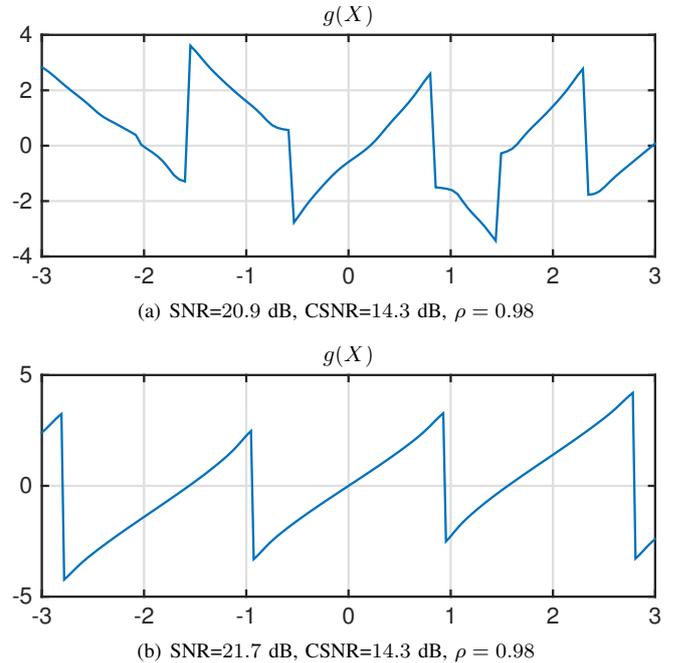

Fig. 4. Two results by NCR for side information setting. In (a) the bins do not have the optimal shape that was obtained by DA and in (b) the discontinuity points are not optimal.

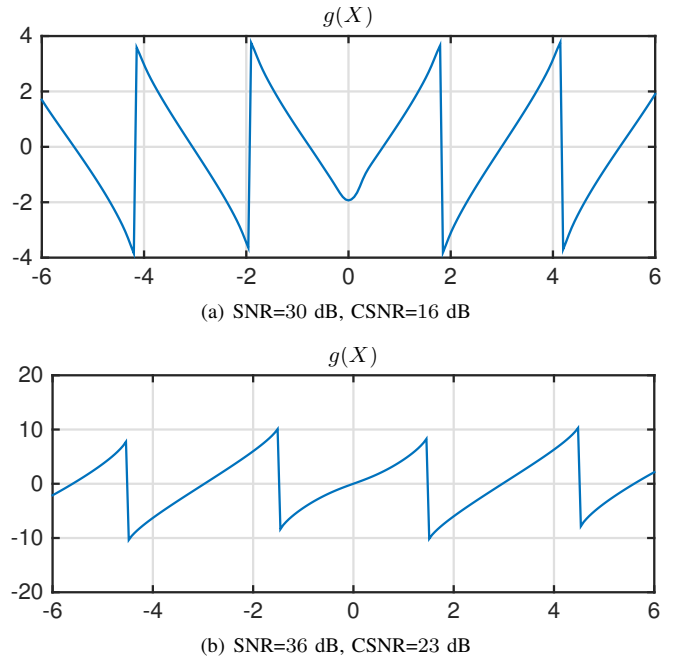

Fig. 5. Example encoder mappings, generated by DA, for Gaussian mixture distribution, side information setting.

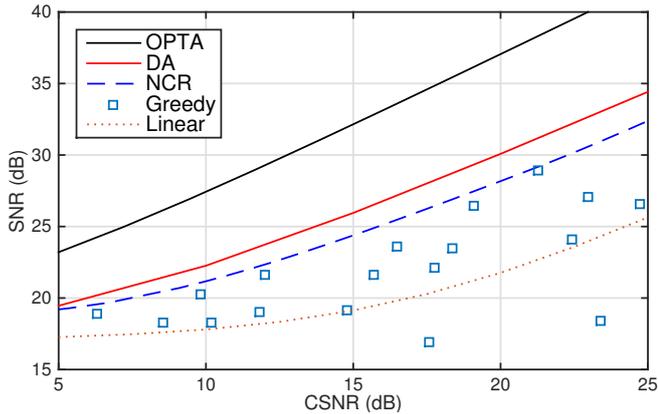

Fig. 6. The performance comparison for the side information setting, the proposed method versus the noisy relaxation (NCR), greedy optimization and the linear mappings. $\rho = 0.99$.

### B. Distributed Coding Setting

In these experiments the sources are jointly Gaussian with unit variance and their correlation coefficient is denoted by $\rho$. We first analyze the case of individual reconstructions, where the cost is as defined in (4). The weighing coefficient $\eta$ in (4) is 1.

The encoding mappings observed are many-to-one, where an example is given in Figure 7a to gain intuition into the workings of these coding schemes. From Figure 7a, where both encoders are plotted together, we see that in different source intervals, one of the mappings is many-to-one while the other one is one-to-one (usually linear). For instance, in the interval $X \in [-0.3, 0.5]$, $g_1$ is approximately linear while $g_2$ is many-to-one. Intuitively, in each of these intervals, one channel is used as side information to reduce the uncertainty about the interval of other source.

Next, we analyze how the channel space is filled. We plot $g_1$ vs. $g_2$ in Figure 7b, which would be the channel space mapping if the two sources were fully correlated ($\rho = 1$). In case of lower correlation, a line widens into a strip (see figures and discussion in [13]), however the plot in Figure 7b is sufficient for demonstration. This mapping has the same characteristics with that of Archimedean spirals used in literature (example plotted in Figure 7c), in the sense that most likely source values are mapped to the area around origin and the mapping continues outwards in a circular fashion, to fill the channel space while preserving transmission power. In fact, spirals are suggested since they have this characteristic. Although our mappings have the same characteristic, they are far different from a spiral.

Spiral-like channel filling may sometimes be sub-optimal. The channel space can be filled in a different way, especially in case of unequal transmission powers. In Figure 8, we provide such mappings where we still see the same characteristics mentioned earlier (both sources acting as side information in different intervals), but the channel space is filled differently. Other examples can be found in literature as well, see, e.g., [12], [13].

In [13], the authors noted that for $0 < \rho < 0.95$, their structured solutions does not improve over linear solutions at high CSNR. We provide an example of non-linear scheme in Figure 9 for $\rho = 0.9$ that improves over linear solution. For lower correlations our method produces linear solutions. Based on these experiments, we reach to a similar conclusion that optimal mappings are non-linear only at high correlation - albeit our method offers non-linear gains over a larger range of $\rho$ values.

Performance comparison of different numerical optimization techniques (DA, NCR and greedy descent with random initialization) for total power allocation case ($\lambda_1 = \lambda_2$) is provided in Figure 10a where we define $\text{SNR} = 10 \log_{10}(2/D)$ (average distortion in dB) and $\text{CSNR} = 10 \log_{10}((P(g_1) + P(g_2))/2)$ (average power in dB). We note that since individual powers are not constrained, different transmission powers are allowed in this comparison for all methods.

We also provide comparison to other coding schemes found in the literature. In [13], authors analyze parametric mappings of two types, spirals and sawtooth mappings, in distributed coding setting and compare to distributed quantizer scheme analyzed in [12]. In their comparison they use same power allocation for both encoders, as opposed to a total power allocation we consider. We therefore obtain solutions that allocate same power to both encoders. In Figure 10b, we provide comparison with our results to the ones reported in [13] for the same setting. As expected, mappings optimized in function space perform better than parametric mappings which only approximately model optimal mappings as demonstrated in Figure 7.

We finally take a look at the function computation problem for which the cost is given in (5). As a test case, we employed the difference function, $\gamma = X_1 - X_2$. Encoder mappings optimized with DA are given in Figure 11a. Both sources are mapped in many-to-one fashion with no way to resolve the uncertainty about the source interval. This is unlike previous mappings, where the uncertainty about source interval is resolved by side information (in the distributed coding case, the other source would act as side information, at least locally). In the case of difference function, the actual values are not needed, thus, both sources are mapped in many-to-one fashion. Nevertheless, the decoder is able to estimate the difference of sources accurately.

We give performance comparison in Table I where $\text{CSNR}_i = 10 \log_{10}(P(g_i))$ for $i = 1, 2$ and $\text{SNR} = 10 \log_{10}(1/D)$. DA achieves 10 dB higher SNR than the linear solution with the same power allocation, whereas the linear solution that achieves the same SNR requires 9 dB more power for each channel. Although the improvements depend on the problem parameters, these results nevertheless demonstrate that there are significant gains in utilizing non-linear encoder functions instead of linear ones. DA performance is better than NCR as well, as the shape of encoders are better optimized as can be seen in comparison in Figure 11.

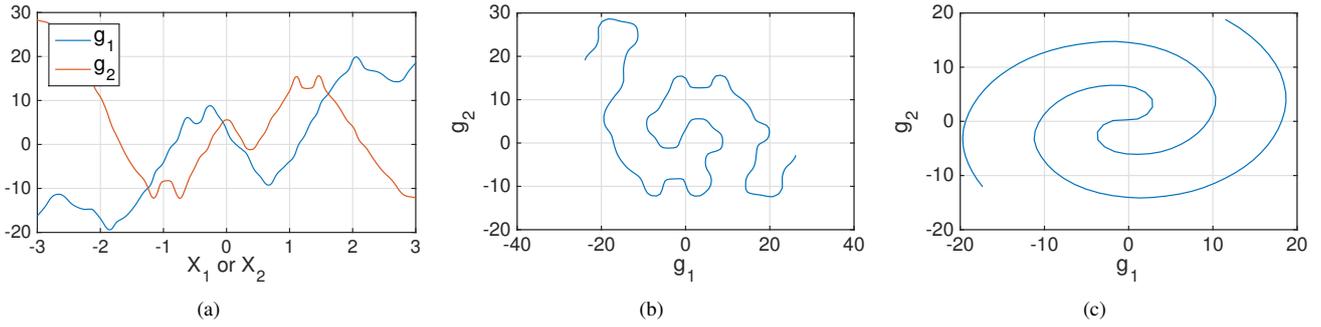

Fig. 7. Example encoding scheme for distributed coding scheme with $\rho = 0.999$. In (a), $g_1$ and $g_2$ are plotted together. In (b) we see how channel space is filled. In (c) a typical Archimedean spiral used in literature is shown.

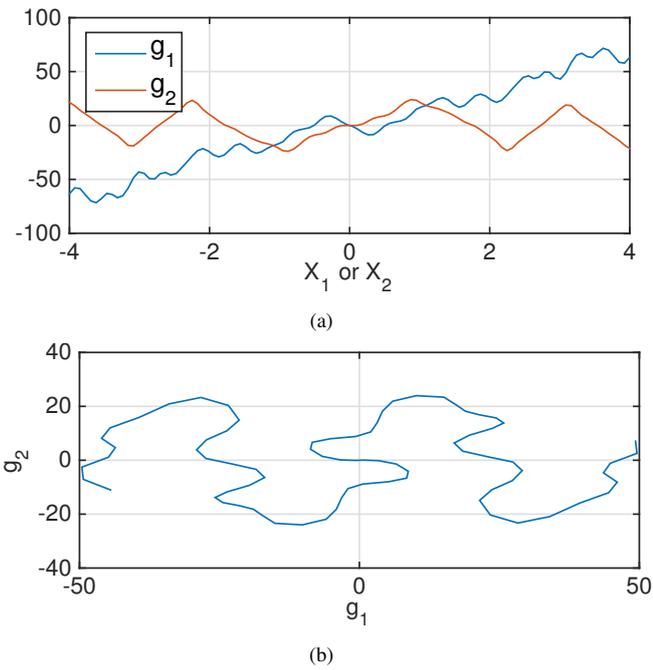

Fig. 8. An example, obtained by DA, with different transmission power constraints on encoders. (a) Both encoders are plotted together. (b) Channel space filling is shown. Although similar characteristics are observed, the channel space is filled differently.

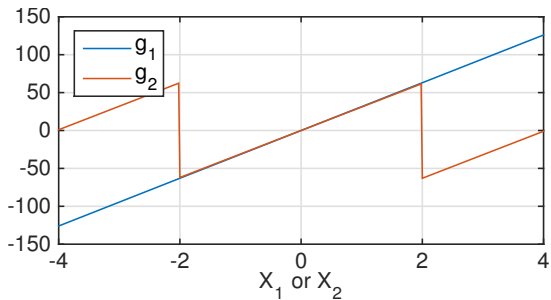

Fig. 9. Non-linear solution that improves over linear for $\rho = 0.9$. CSNR = 29 dB, SNR = 29.82 dB. Linear solution at same CSNR achieves SNR = 29.60 dB.

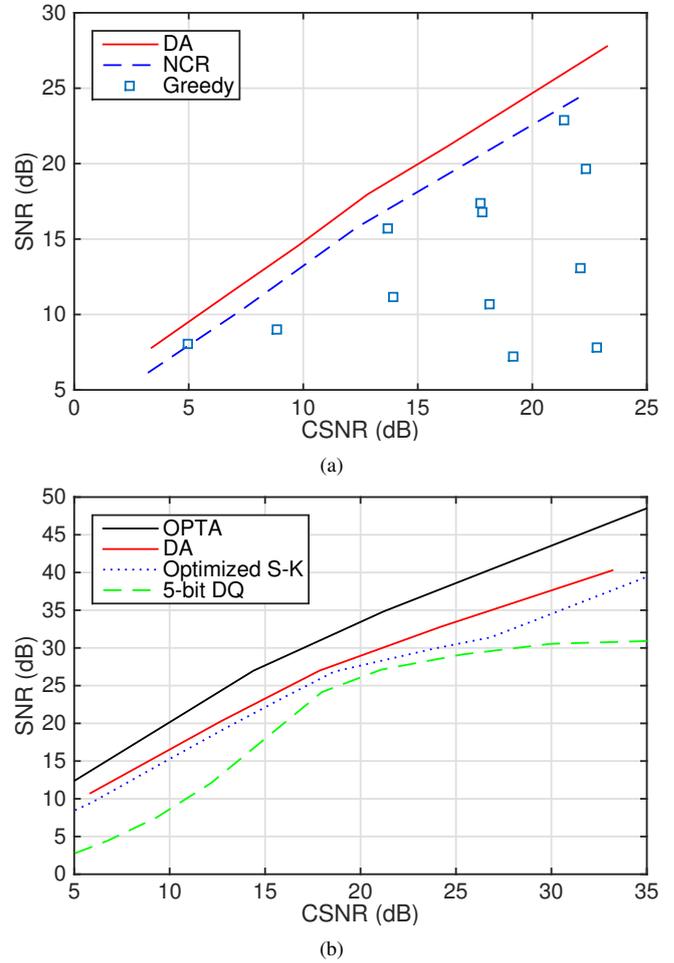

Fig. 10. (a) Performance comparison of different numerical optimization methods for distributed coding setting with the constraint on total transmission power. $\rho = 0.99$. (b) Performance comparison for distributed coding setting with other approached found in literature. Optimized S-K refers to performance of structured mappings in [13] (spirals and sawtooth mappings) and 5-bit DQ is from [12]. 5-bit DQ is optimized for 18 dB CSNR. $\rho = 0.999$.

TABLE I
PERFORMANCE OF OBTAINED MAPPINGS FOR DIFFERENCE FUNCTION

| Method | $\text{CSNR}_1$ (dB) | $\text{CSNR}_2$ (dB) | SNR (dB) |
|---|---|---|---|
| DA | 19.9 | 21.4 | 27.3 |
| Linear-1 | 19.9 | 21.4 | 17.0 |
| Linear-2 | 28.9 | 30.4 | 27.2 |
| NCR | 19.9 | 21.5 | 24.0 |

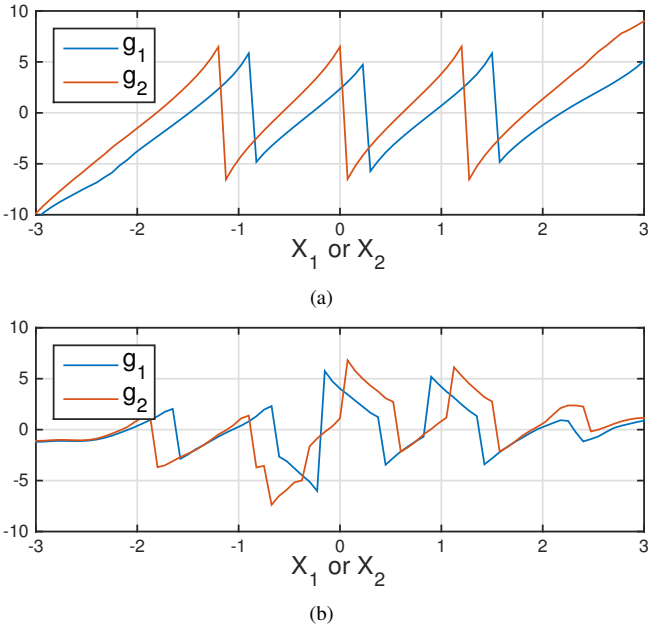

Fig. 11. Example solutions obtained for function computation problem, where $\gamma = X_1 - X_2$. (a) DA result (b) NCR result. CSNR and SNR values are in Table I.

## VI. CONCLUSIONS

In this paper, we studied the problem of finding globally optimal encoder and decoder pairs in zero delay source-channel coding, focusing on two basic network settings. Since the cost surface is riddled with local optima, we developed a method based on the deterministic annealing to approach global optimality. The numerical results show that, by using carefully optimized non-linear (and in many cases many-to-one) mappings, significant gains can be obtained over linear solutions, which are optimal in point-to-point settings (for the specific case of Gaussians under MSE). Simulation results demonstrate the performance of the proposed algorithm, which consistently outperform greedy optimization methods and noisy channel relaxation, as well as the previous approaches found in literature.

## APPENDIX

**Proof of Theorem 1**: We write the Lagrangian cost in (25) as

$$F = \sum_k \int_{\boldsymbol{x}} J_k(\boldsymbol{x}) p(k|\boldsymbol{x}) f_{\boldsymbol{X}}(\boldsymbol{x}) \mathrm{d}\boldsymbol{x} - \lambda P_E$$
$$+ T \sum_k \int_{\boldsymbol{x}} p(k|\boldsymbol{x}) \log p(k|\boldsymbol{x}) f_{\boldsymbol{X}}(\boldsymbol{x}) \mathrm{d}\boldsymbol{x} + T H_0, \quad (39)$$

where $J_k(\boldsymbol{x})$ is given in (27). From (39) it can be seen that $F$ is convex in $p(k|\boldsymbol{x})$, since first term is linear and second term is convex in $p(k|\boldsymbol{x})$. To find the minimum, we set $\nabla_{p(k|\boldsymbol{x})} F = 0$:

$$J_k(\boldsymbol{x}) + T \log p(k|\boldsymbol{x}) + T = 0, \quad (40)$$

which yields

$$p(k|\boldsymbol{x}) = C e^{-(J_k(\boldsymbol{x})-T)/T}. \quad (41)$$

The normalizing factor $C$ is to ensure that

$$\sum_k p(k|\boldsymbol{x}) = 1. \quad (42)$$

Plugging (41) in (42), we have

$$C = \frac{1}{\sum_{k'} e^{-(J_{k'}(\boldsymbol{x})-T)/T}}. \quad (43)$$

Plugging (43) in (41) yields (26).